\documentclass {article}

\usepackage{amssymb}
\usepackage{amsmath}
\usepackage{float}
\usepackage{rotating}
\usepackage{epsfig}

\begin{document}

\title {Statistical physics applied to stone-age civilization}
\maketitle
\author {M.A. Sumour $^{1} $, \and M.A. Radwan$^{1} $, \and M.M. Shabat$^{2} $,  \and Ali H. El-Astal$^{1} $}\\
\\
$^1$ Physics Department, Al-Aqsa University, P.O.4051, Gaza,
Gaza Strip, Palestinian Authority,
e-mail: msumoor@yahoo.com , ma.radwan@alaqsa.edu.ps , 
a\_elastal@yahoo.com .\\
$^2$ Physics Department, Islamic University, P.O.108, Gaza, Gaza Strip, Palestinian Authority, 
e-mail : shabat@iugaza.edu.ps

\section*{Abstract}

About 45,000 years ago, symbolic and technological complexity of human artefacts
increased drastically. Computer simulations of Powell, Shennan and Thomas (2009)
explained it through an increase of the population density, facilitating the
spread of information about useful innovations. We simplify this demographic
model and make it more similar to standard physics models. For this purpose, 
we assume that bands (extended families) of stone-age humans were distributed randomly
 on a square lattice such that each lattice site is randomly occupied
with probability $p$ and empty with probability $1-p$. Information spreads randomly from 
an occupied site to one of its occupied neighbours.\\
If we wait long enough, information spreads from one side of 
the lattice to the opposite site if and only if $p$ is larger than
 the percolation threshold; this process was called "ant in the labyrinth" by de Gennes 1976.
 We modify it by giving the diffusing information a finite lifetime, which shifts the threshold upwards

\section*{Introduction:}

Modern humans presumably originated about 200,000 years ago in East Africa.
They emigrated to 150 km north of Gaza about 100,000 years ago (Skhul near
Haifa) but died out there again. Then about 50,000 years ago they again emigrated 
from Africa, this time more successfully, but presumably again via 
Gaza and Haifa.\\
According to [1], demographic factors  can explain geographic variation in the 
timing of the first appearance of modern behaviour without invoking increased 
cognitive capacity.\\
Now we try to reproduce this demographic effect by applying standard percolation theory [2].
 Only for densities p above the percolation threshold  
$(p_c= 0.593$ 
on infinite square lattices) can information travel from one side of the lattice, 
via random jumps to occupied nearest-neighbour sites, up to the opposite side of 
the lattice. This type of random walk on random square lattices was discussed a 
lot about three decades ago, but mostly for large systems and long times close to 
the percolation threshold. We will show that for times below 10,000 jump attempts 
even for quite small lattices, one needs occupation probabilities  far above the 
percolation threshold ($p_c= 0.593$) in order to transfer information from 
one side of the lattice to the opposite site.

\section*{Data and Simulations:}

First we have occupy a square lattice of size $Lx * Ly$ randomly with
probabilities $p$ for occupied and $1-p$ for empty, clusters are groups of occupied 
neighbouring sites. Then  we put $N$ diffusing particles randomly onto occupied 
sites. Then we  let each particle at each time step try to move into one of the 
four possible directions; if that neighbour is occupied the particle (which may 
represent a teacher of new techniques) moves there, if it is empty the particle 
stays at its old place.\\
 In this way, $N$ particles diffuse on a randomly occupied lattice, using only the 
 occupied lattice sites. The occupied sites can be settlements of humans, and the 
 diffusing particles can be visitors spreading information on new techniques.\\
We start our simulations with  for different $Lx=10,20,30,40,50,100,200,300$. With 
$Ly=100,1000$ ,  with different probability$=0.60,0.70,0.80,0.90$. We check after 
which time, i.e. after how many jump attempts, the particle diffuses across the 
lattice in x-direction.\\
And we have nine samples $(N=9)$ ($N$ diffusing particles on the same lattice),  from 
which the median is defined such that four times are larger and four times are 
smaller than the median time, shown in Fig. 1.

\begin{figure}[H]
  \centerline {
    \epsfig{figure=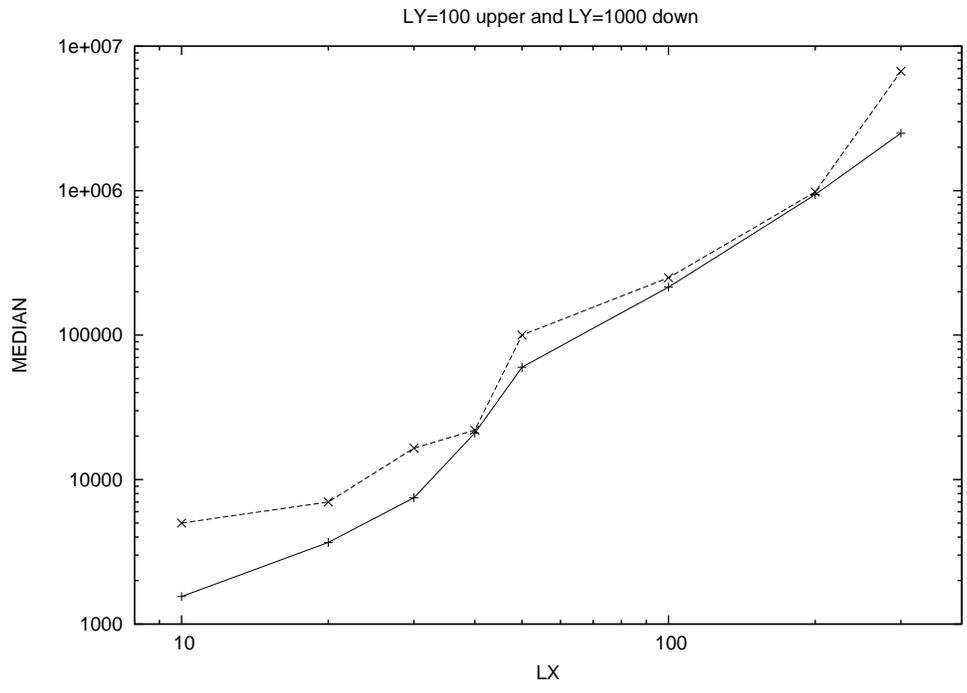, height=5.0in , angle=270}
     }
      \caption{Median time versus $Lx$ for $Ly=100$ upper line and  
      the down line for $Ly=1000$ , iseed=1,at $p = 0.64$ .}
       \label{Fig1}
\end{figure} 

Then we complete our simulations for three probabilities $=0.70,0.80,0.90$ and take 
the median time, $Ly=100$ with random seed number iseed=1, so we get figure (2):

\begin{figure}[H]
  \centerline {
    \epsfig{figure=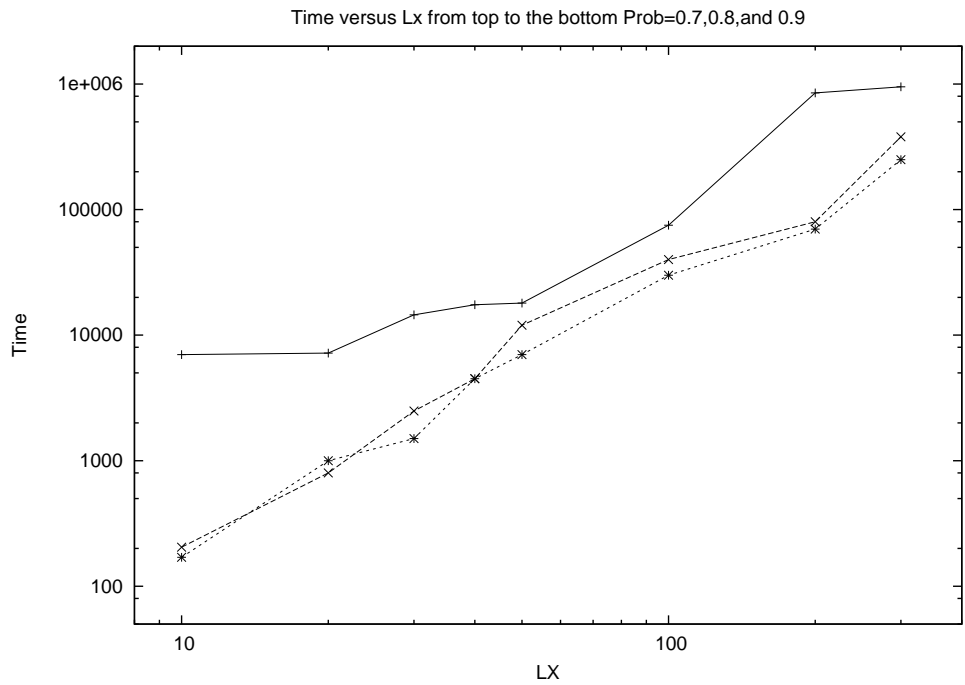, height=5.0in , angle=270}
     }
      \caption{ Median time versus $Lx$ for $Ly=100$ from top to 
      bottom $prob.= 0.70,0.80,{\&} 0.90$ with iseed=1.}
       \label{Fig2}
\end{figure}

Figure (3) change the random number as 
iseed=1,2,3 for one sample of probability =0.70, and then we take the 
average median for each iseed number to get figure (3): 

\begin{figure}[H]
\centerline {
    \epsfig{figure=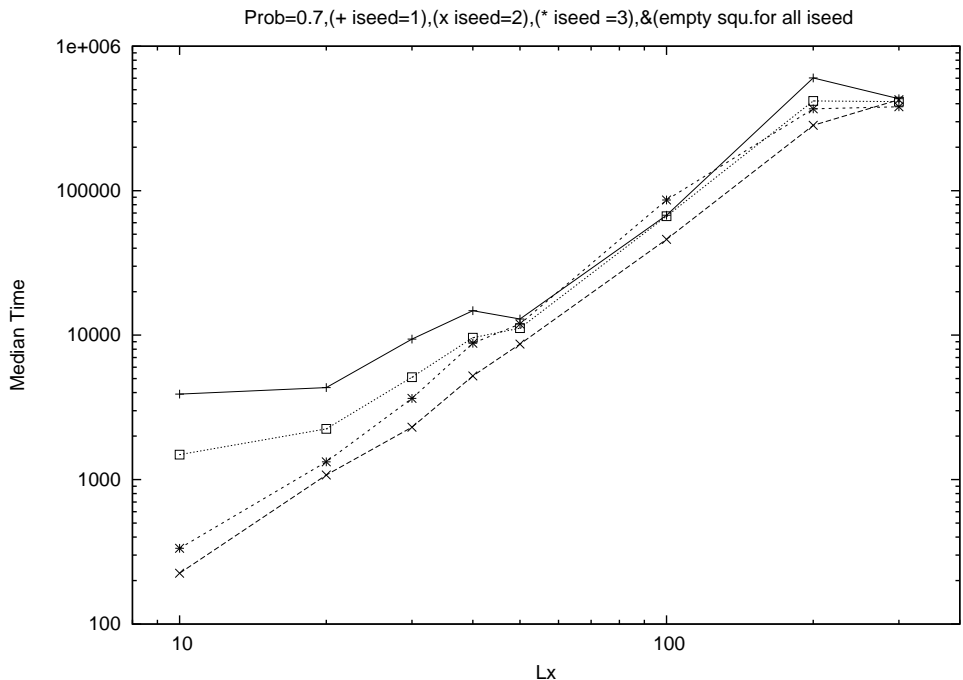, height=5.0in , angle=270}
     }
     \caption{Median time for $p = 0.7$ fluctuations are reduced
      by averaging over three random lattices for $Ly=100$ .}
      \label{Fig3}
\end{figure}

\section*{Conclusions:}

If we identify one time unit with one day, 
times above 10,000 are unrealistic for a single messenger of new techniques. 
Thus only rather large occupation probabilities closer to unity than to $p_c$ 
allow the random spread over dozens of distances between human bands. Future 
simulations might facilitate information spread by allowing these bands to 
move randomly on the lattice (annealed instead of quenched disorder), as was 
appropriate before the Neolithicum with agriculture.

\section*{Acknowledgements:}

 The authors would like to thank Prof. Stauffer 
for many valuable suggestions, fruitful discussions and constructive advice 
during the development of this work.

\section*{References:}

[1] Adam Powell, Stephen Shennan, and Mark G. 
Thomas, Science 324, 1298 (2009).
\newline 
[2]Dietrich Stauffer and Amnon Aharony, 
Introduction to Percolation Theory, Taylor and Francis, London 1994
 (2nd printing of 2nd edition)

\end{document}